\documentclass[a4paper,byrevtex]{revtex4}
\usepackage{dcolumn}
\usepackage{amsmath}
\usepackage{bm}
\usepackage{graphicx}

\begin{document}

\title{$\bm c$ at the belfry}

\author{Claude \surname{Semay}}
\email[E-mail: ]{claude.semay@umons.ac.be}
\affiliation{Service de Physique Nucl\'{e}aire et Subnucl\'{e}aire,
UMONS Research Institute for Complex Systems,
Universit\'{e} de Mons,
Place du Parc 20, 7000 Mons, Belgium}
\author{Francesco \surname{Lo Bue}}
\email[E-mail: ]{francesco.lobue@umons.ac.be}
\author{Soizic \surname{M\'elin}}
\email[E-mail: ]{soizic.melin@umons.ac.be}
\affiliation{Service Sciences et Techniques au Carr\'{e},
Universit\'{e} de Mons,
Place du Parc 20, 7000 Mons, Belgium}
\author{Francis \surname{Michel}}\thanks{Retired professor of the Universit\'{e} de Mons}

\date{\today}
\begin{abstract}
In 1849, Hippolyte Fizeau determined the speed of light in a famous experiment. The idea was to measure the time taken for a pulse of light to travel between an intense light source and a mirror about 8 km away. A rotating cogwheel with 720 notches, that could be rotated at a variable speed, was used to chop the light beam and determine the flight time. In 2017, physicists and technicians of the University of Mons in Belgium reproduced the experiment with modern devices to allow members of the public to measure the speed of light themselves. The light source used was a low power laser, and the cogwheel was replaced by an electrically driven chopper, but the general spirit of Fizeau's experiment was preserved. The exhibition was organised in the belfry of Mons, a baroque-style building classified as a UNESCO World Heritage site. The solutions found for the main problems encountered are presented here to help colleagues intending to reproduce the experiment. 
\end{abstract}




\maketitle

\section{The genesis}
\label{sec:genesis}

On the occasion of World Year of Physics in 2005, physicists and technicians at the University of Mons (UMONS) in Belgium duplicated the famous Foucault pendulum experiment in the town's Saint Waltrude Collegiate Church \cite{UMONS2015}. The response of the public was impressive and many positive comments were received from visitors. Following a suggestion by one of the authors (FM), the same team decided to start a new project: to repeat Hippolyte Fizeau's famous 1849 experiment to measure the speed of light \cite{fizeau,frercks,lequeux,hughes}. 

At the same time, physicists from the \emph{Observatoire de Paris} were developing their own version of the experiment in the framework of the exhibition ``$c$ \`a Paris"\footnote{``$c$ \`a Paris" is a pun playing with the homophony in French between the pronunciation of the letter ``c", to designate the speed of light, and the locution ``c'est" -- ``C'est \`a Paris" meaning ``It is in Paris".}. They used a 5 watt laser, which is dangerous to manipulate with people nearby. Moreover, the flight time between the emitter at the \emph{Observatoire} and the reflector at Montmartre was determined by using a complex electro-optic device, which is difficult for non-experts to understand. 

Our purpose was very different. First, we wanted to preserve the general idea of the original experiment, but using modern devices. Reproducing Fizeau's original apparatus would have been very difficult, extremely expensive, and not very interesting from a pedagogical point of view. Indeed, our second objective was to explain the principles of the experiment to visitors and allow them to measure the speed of light by themselves.

In the 1849 experiment, Fizeau measured the time taken for a pulse of light to travel between an intense light source and a mirror about 8 km away. Details of the apparatus and the procedure can be found in a paper by Frercks \cite{frercks} who duplicated the experiment to investigate the techniques used by Fizeau. They are also presented in many physics textbooks, in books dedicated to the life of famous physicists \cite{lequeux,hughes}, and even in docufiction \cite{canalu}. Here we present the main features of the Fizeau's experiment to highlight the differences between the original experiment and our modern version. In the original experiment, a heliostat or a Drummond light was used as the light source, and a rotating cogwheel with 720 notches that could be rotated at a variable speed was used to chop the light beam. When the wheel is at rest, a beam of light passing through a notch is reflected by a mirror and comes back to the observer's eye through the same notch. If the cogwheel is rotated quickly enough, then the notch through which the light passed can be replaced by the adjacent tooth, which thus blocks the reflected light on its return path. From the rotational speed of the wheel and the distance between the wheel and the mirror, Fizeau was able to deduce a value of about 315~000~km/s for the speed of light. 

We kept exactly the same principle but with modern devices. Our light source was a low power laser of 5~mW, which is safe for members of the public to use. The cogwheel was replaced by an electrically driven chopper, which has a moderate cost and is very easy to operate. The reflected light was filmed by a CCD camera and sent to a PC to help visitors taking the measurement. 

In the following, we will not explain in full detail and in chronological order all the steps in the development of the experiment\footnote{The full story (in French) is detailed in \cite{UMONS2017}. Pictures from the development of the experiment and a movie showing two extinctions during a measurement are available at \texttt{http://scitech2.umons.ac.be/c-at-the-belfry/}.}, but instead just present solutions to the main problems faced. Between the initial idea and the exhibition, 12 years elapsed. This was not only due to the difficulty of the task, but also because members of the team only worked episodically on the experiment. Moreover, it was necessary to find agencies and convince them to provide funding in order to offer a good quality exhibition to the public. At last, one other significant reason for the length of time needed will be explained in the next section. 

\section{The sites}
\label{sec:sites}

Our first problem was to find two sites to host the emitter and the reflector. We needed two locations, elevated above their surroundings so that the laser beam could not be intercepted accidentally. They had to be separated by several kilometres in order to ensure the feasibility of the measurement. As was the case of our duplication of the Foucault pendulum experiment, we wanted places full of history to offer a beautiful setting for visitors. The main optical device had to be located in a room capable of accommodating about thirty visitors, with sufficient additional space to explain the experiment with teaching material. The reflector had to be located in a safe place sheltered from any disturbances. 

One of the key elements of the apparatus was the simplicity of the reflector: we used a simple panel of one square metre covered with retroreflective sheeting, like that used for road signs, provided by the firm 3M\,\texttrademark. In this way, the light is reflected back to its source with minimum loss, as the panel is roughly perpendicular to the laser beam. Once installed, no further maintenance or monitoring was necessary. 

\begin{figure}[htb]
\includegraphics[width=8.54cm,height=5cm]{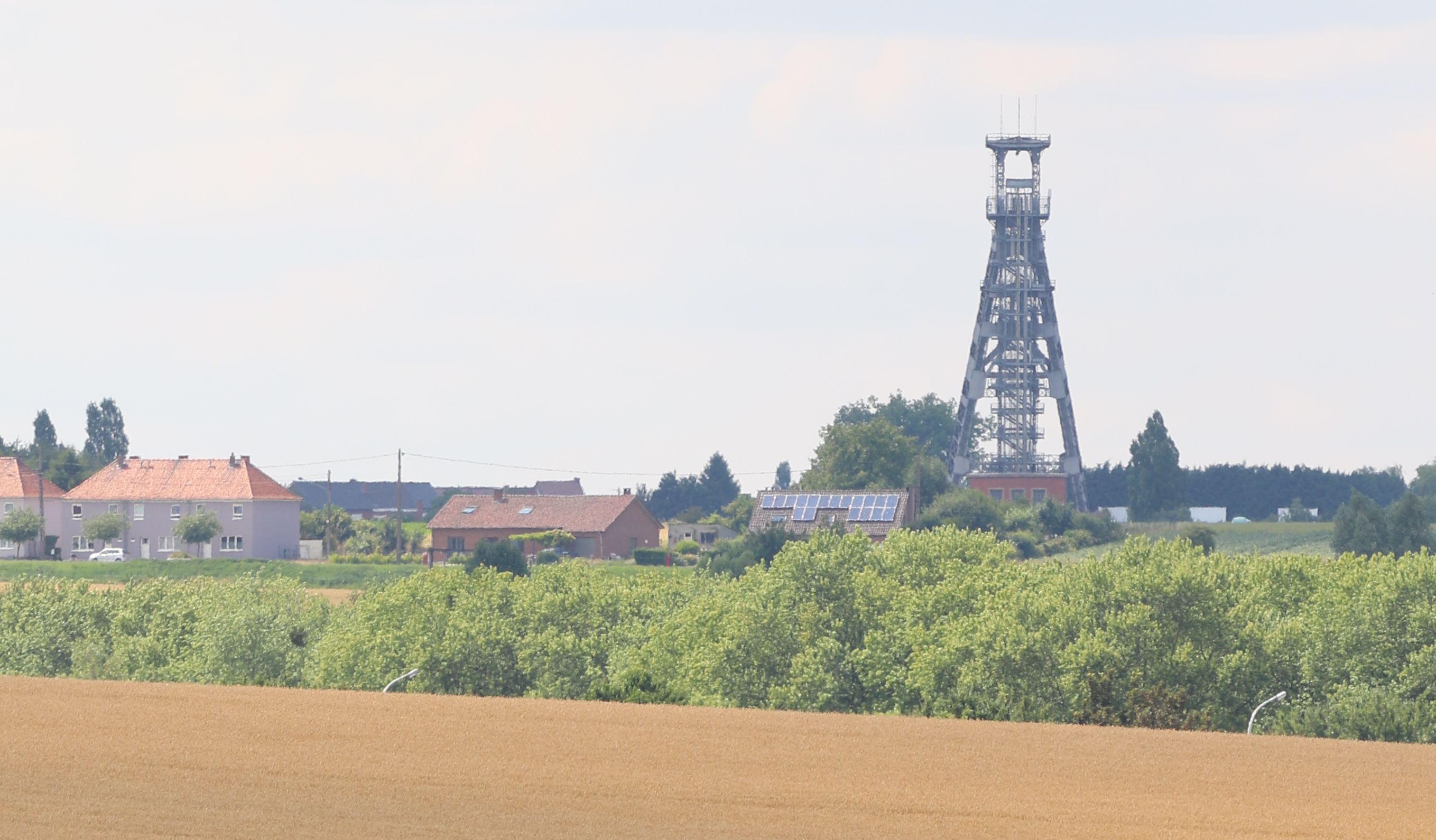} 
\includegraphics[width=8.29cm,height=5cm]{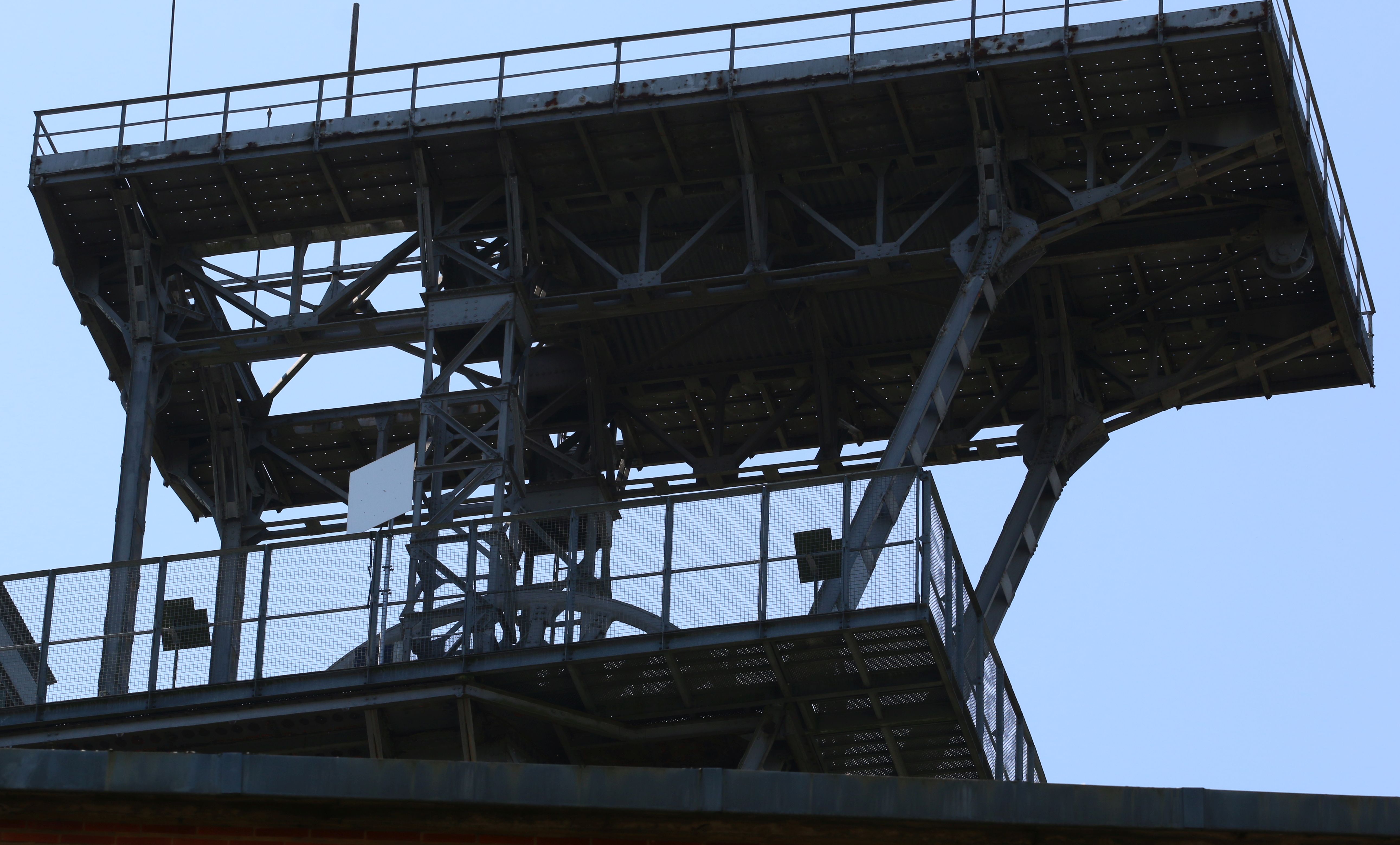} 
\caption{Left: the headframe of the Pass on the horizon. Right: the small white panel just below the roof of the headframe is the reflector.}
\label{fig1}
\end{figure}

\begin{figure}[htb]
\includegraphics[width=9.98cm,height=5cm]{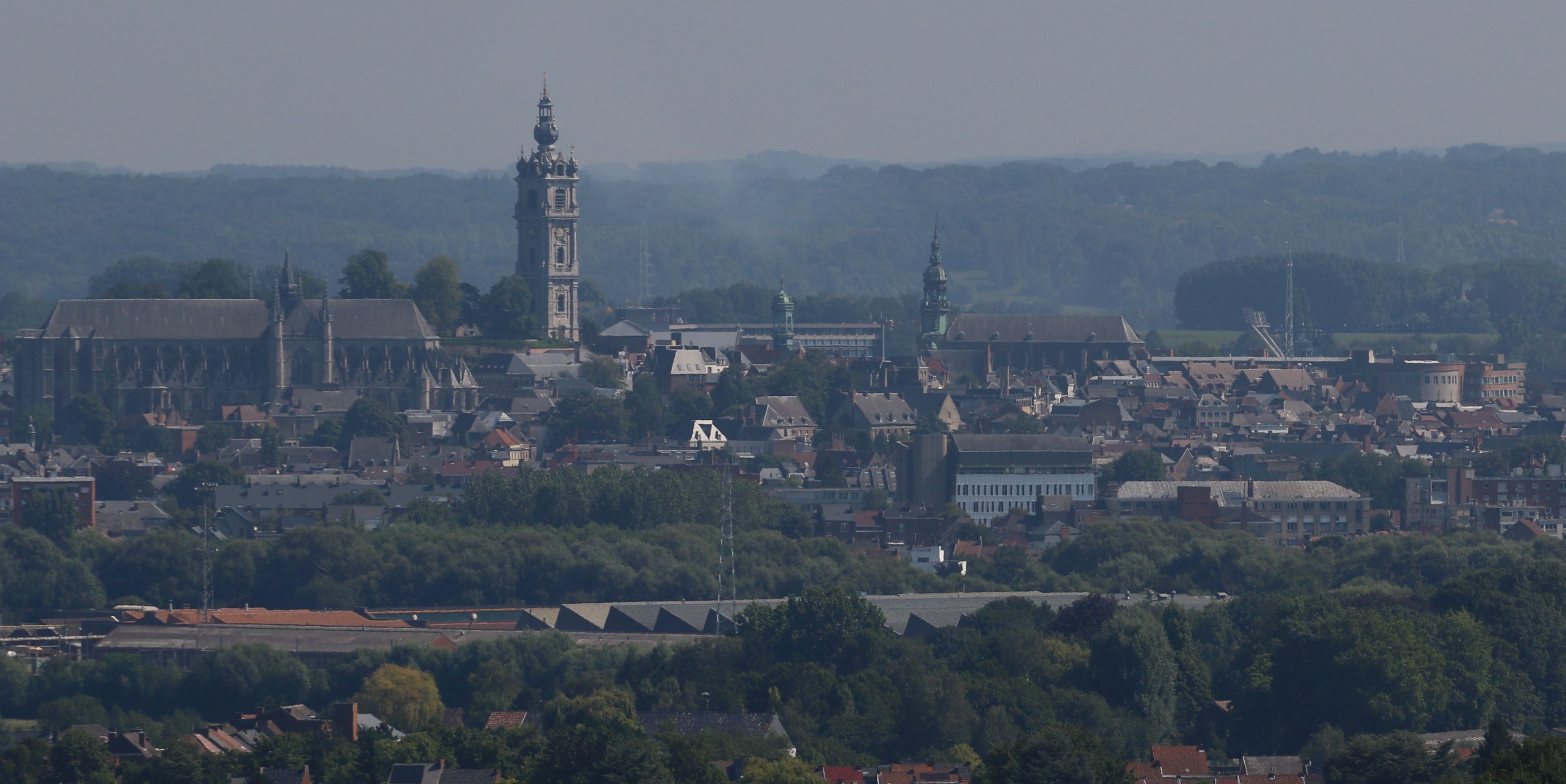} 
\includegraphics[width=2.81cm,height=5cm]{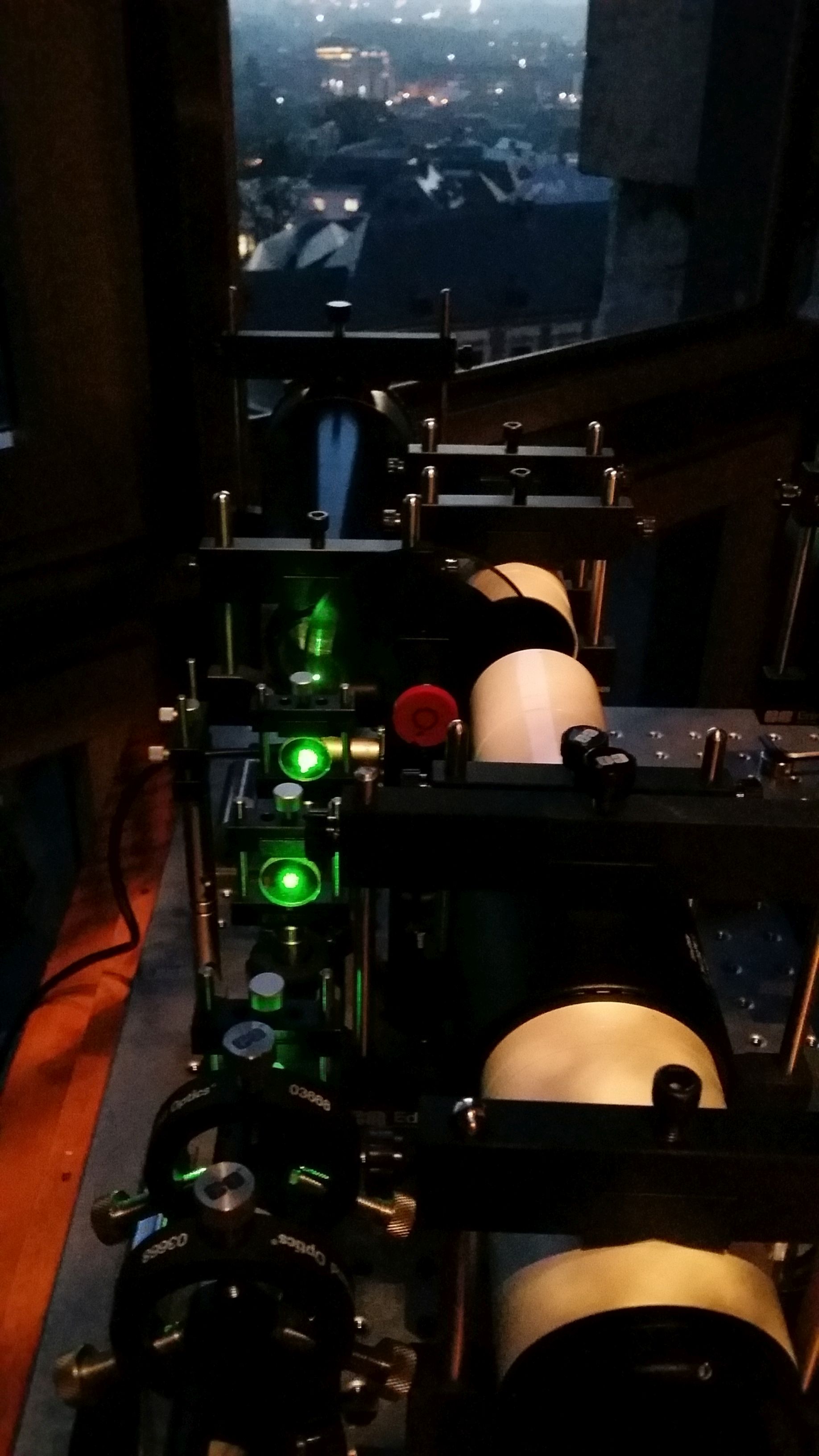} 
\caption{Left: the belfry dominating the skyline of the city of Mons, as seen from the top of the headframe of the Pass (picture taken with a telephoto lens). Right: the optical device, on the third floor of the belfry, in operation during the evening.}
\label{fig2}
\end{figure}

After examining several possibilities, we chose to place the reflector on the top of a headframe located at the `Pass' (Parc d'aventures scientifiques -- Scientific adventure park) (see fig.~\ref{fig1}, \texttt{http://www.pass.be}), an establishment for promoting scientific culture, located on the site of an old coal mine. The Mons belfry, a baroque-style building classified as a UNESCO World Heritage site (see fig.~\ref{fig2}, \texttt{http://www.beffroi.mons.be/}), was chosen as the place to host the optical device. This is the origin of the title of the exhibition (and of this paper) ``$c$ at the belfry"\footnote{We would like to thank our colleagues from the \emph{Observatoire de Paris} who allowed us to adapt the title of their exhibition.}, with the subtitle ``You are invited to come and measure the speed of light" \cite{UMONS2017}. These two locations, which are both strongly connected to the history of the city and to the promotion of popular science, are about 5.4~km apart. This distance is long enough to ensure the possibility of correct measurements within the range of our chopper, and short enough to make the pointing of the laser not too difficult. As a precise determination of the distance was not sought, it was simply determined by GPS position reports obtained from \emph{Google Earth}. When the project was initiated, the belfry was closed for restoration and these restoration works were estimated to last several years. Finally, with the restoration accumulating long delays, a brand-new belfry only became accessible to the public and our team in July 2015. 

\section{The optical device}
\label{sec:device}

\begin{figure}[htb]
\includegraphics[width=9.02cm,height=5cm]{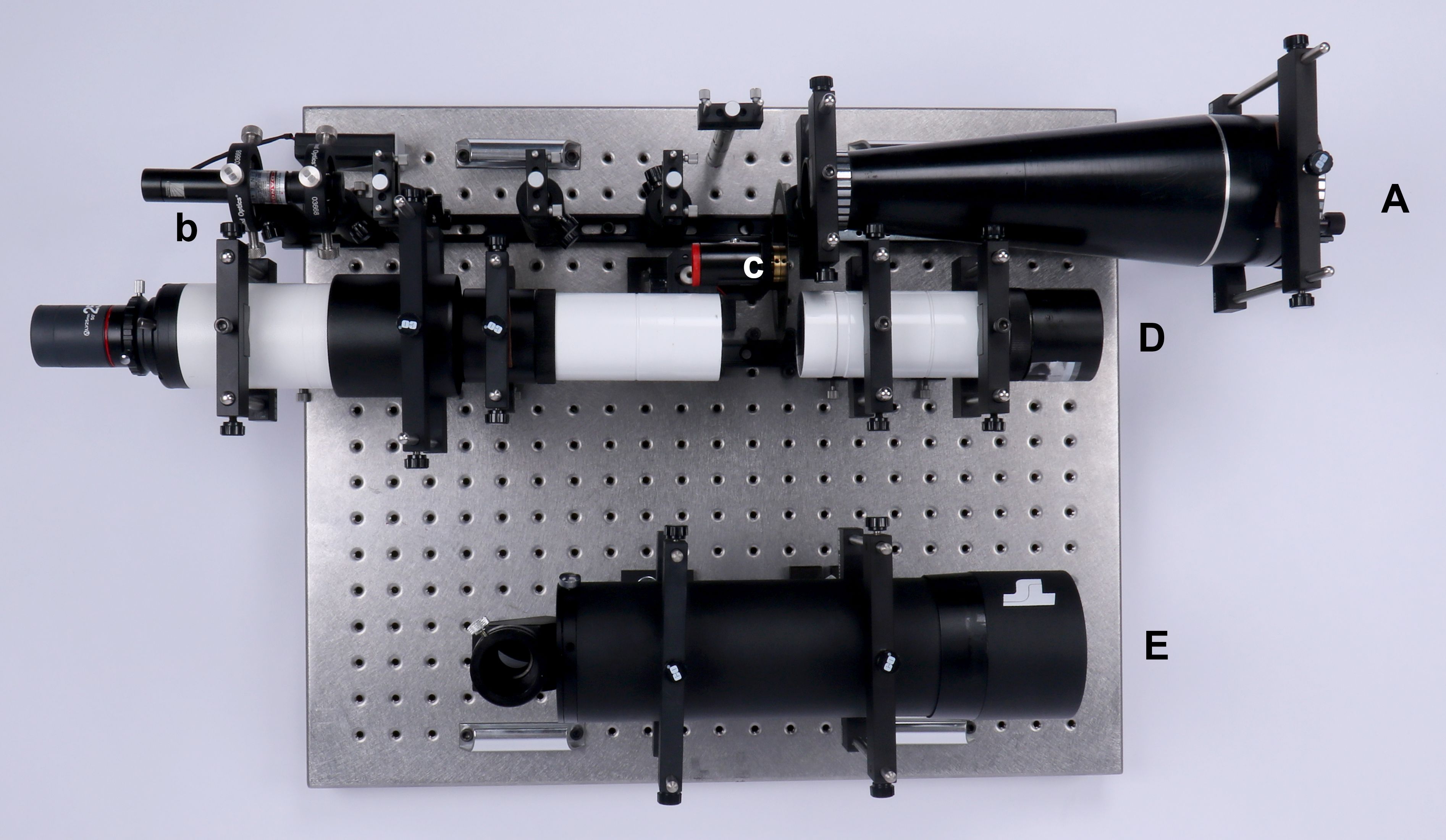} 
\includegraphics[width=6.51cm,height=5cm]{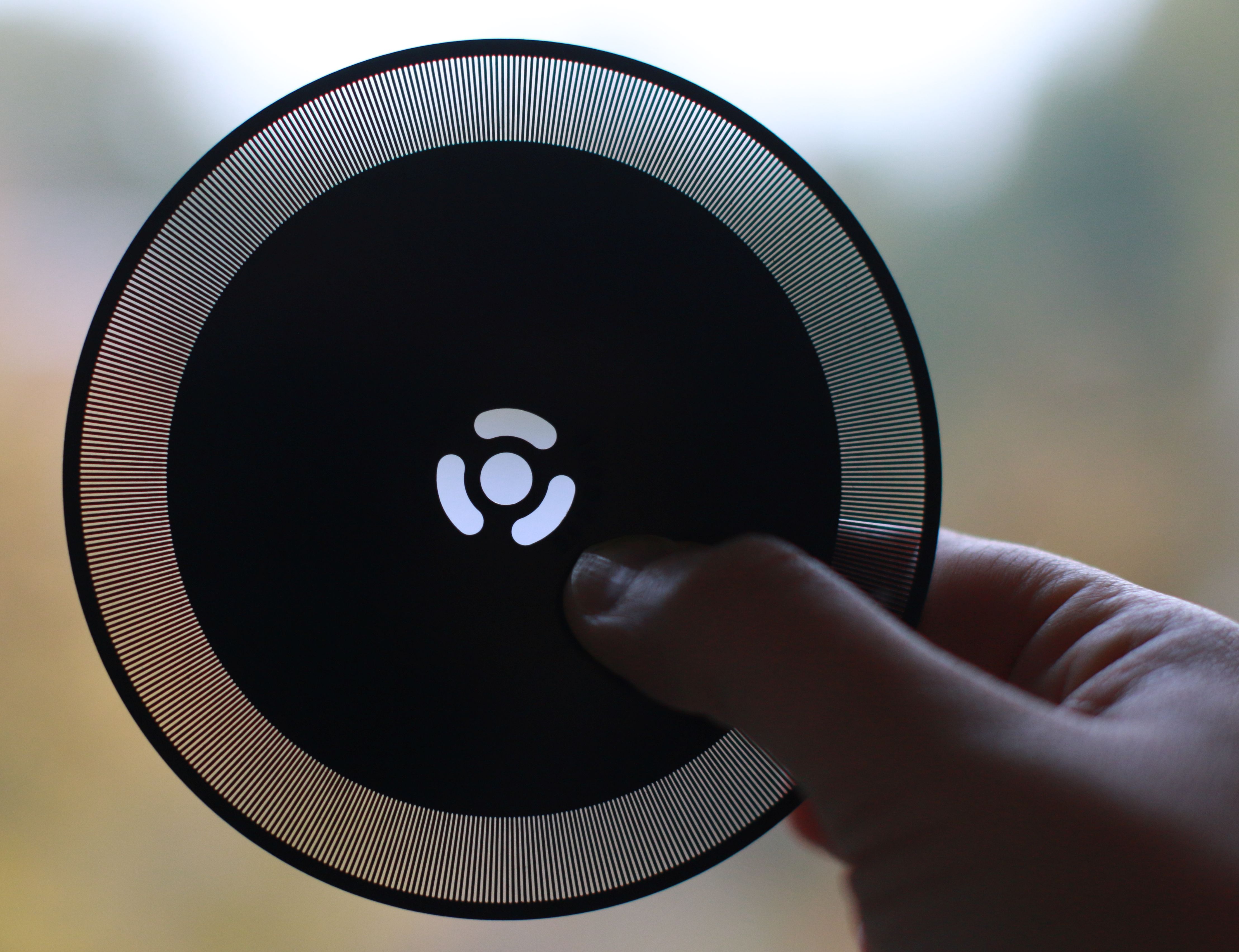} 
\caption{Left: the optical device in the belfry includes the firing line (A), the laser (b), the chopper wheel and its motor (c), the return line (D) and the sight line (E). Right: the core of the optical device is the chopper wheel with 445 slots and a maximum rotation speed of 100 rps. All the apparatus was firmly fixed on to heavy table (one inch between two neighbouring fixation holes).}
\label{fig3}
\end{figure}

All the apparatus of the optical device (laser, lenses, wheel) located in the belfry was firmly fixed on a heavy breadboard laboratory table, which was itself fixed to a rigid tripod in order to reduce vibrations (see fig.~\ref{fig3}). The chosen laser had a low power of 5 mW. It is not dangerous for the public, but is bright enough to allow measurements in broad daylight in various meteorological conditions\footnote{Belgium is not famed for its sunny weather!}. The sun only disturbed the experiment when it was close to the reflector, because it then became a source of unwanted light, spoiling the measurements. Fog prevented certain measurements, but moderate rain did not. A green 532~nm laser was chosen, since the eye is particularly sensitive to this colour, though this was not a crucial point of the experiment. The laser beam passed through a first group of lenses on the firing line, which focused the light on the chopper wheel. The beam was then narrow enough to either pass unaltered through a slot, or to be totally blocked by a blade on the wheel. This focal point was located on the focus of a beam expander with a diameter of 10~cm; this made the beam reasonably parallel before it reached the distant reflector. With this setting the width of the beam on the reflector was estimated, by direct observation, to be about 30~cm. The reflected light passed through a third lens on the return line, which focused it again on the wheel, in order that the beam could pass unaltered through a slot or be totally blocked by the wheel. Finally, a fourth lens made the beam parallel before being sent to the CCD camera. The role of the sight line was to facilitate laser pointing. All the lens groupings could be finely adjusted using micrometre screws. 

The core of the optical device is the chopper wheel, which is 102 mm in diameter and has 445 slots (see fig.~\ref{fig3}). Its maximum rotation speed is 100 rps. At this speed, the time for the replacement of a slot by the adjacent blade is $1.1\times 10^{-5}$~s. With our device, the source light and reflected lights do not pass through the same slot of the wheel, like in Fizeau's experiment, but through diametrically opposite points of the wheel. This helped to reduce the problems of unwanted reflections. As such, at rest, a beam passing through the firing line does not necessarily pass unaltered through the return line. However, the degree of transmission can be finely adjusted using micrometre screws on the return line. In practice, to take a measurement, the lens groupings of the return line are first adjusted at rest, so that a near maximum of light reaches the camera. By increasing the rotation speed of the wheel, the reflected light is progressively blocked, until a minimum is observed. With further increases of the rotation speed, it is easy to understand that one can, in principle, observe successive maxima and minima of the signal. As the time of a two-way journey of the light between the laser and the reflector is $3.6\times 10^{-5}$~s, our device makes the observation of only two successive minima possible. A green interference filter, corresponding to the wavelength of the laser light, with a bandwidth of only 2~nm, was placed in front of the camera to eliminate a maximum of unwanted light and enhance the contrast. The sight line was also equipped with the same kind of filter.

To analyse the quantity of light received by the camera, we used some simple software, \emph{RSpec} (\texttt{https://www.rspec-astro.com/}), which is popular in the world of amateur astronomy. With \emph{RSpec}, a horizontal band in the image provided by the CCD camera can be selected to measure the intensity of light it contains. A minimum of light is reached when the peak intensity is reduced to the level of the background noise, or to a small residual peak when the sky is very bright. As it is difficult to precisely determine the location of a maximum of intensity, we decided to have our measurements rely on the detection of two successive minima of light. Assuming that these are detected at chopping frequencies $f_1$ and $f_2 > f_1$, the time for a two-way journey in seconds is given by $1/(f_2 - f_1)$, where the frequencies are given in Hz. Finally, the measured speed of light $c_m$, in m/s, is given by: 
\begin{equation*}
c_m = 2 d (f_2 - f_1),
\end{equation*} 
where $d=$~5~368~m is the distance between the optical device and the reflector. As stressed above, the purpose of the experiment was mainly pedagogical. Therefore, measurement errors in the determination of the speed of light were only being crudely estimated . By testing the accuracy of the GPS position reports, the error in $d$ was estimated to be about $\pm$5~m. The rotation frequency of the chopper was adjustable by 10~Hz increments. However, the minima could not be located to the naked eye with such a precision. Therefore, the errors in $f_1$ and $f_2$ were estimated to be about $\pm$500~Hz. Typical values for the extinction frequencies were, e.~g., 10~000~Hz and 38~000~Hz. The error in $c_m$ was therefore dominated by the error in frequencies, and expected to be of the order of $\pm$4\%. 

\section{The public experiments}
\label{sec:exp}

\begin{figure}[htb]
\includegraphics[width=6.14cm,height=5cm]{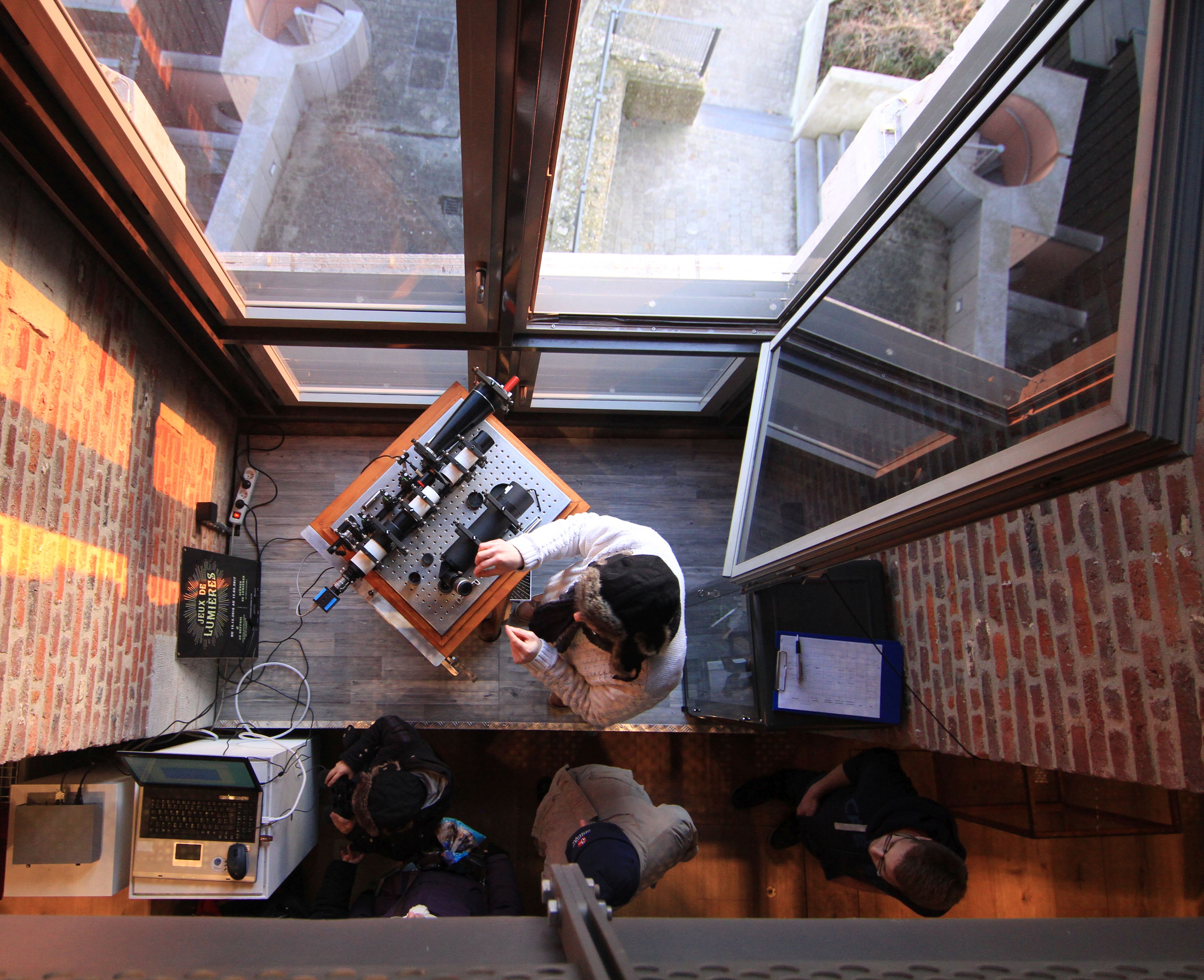} 
\includegraphics[width=6.73cm,height=5cm]{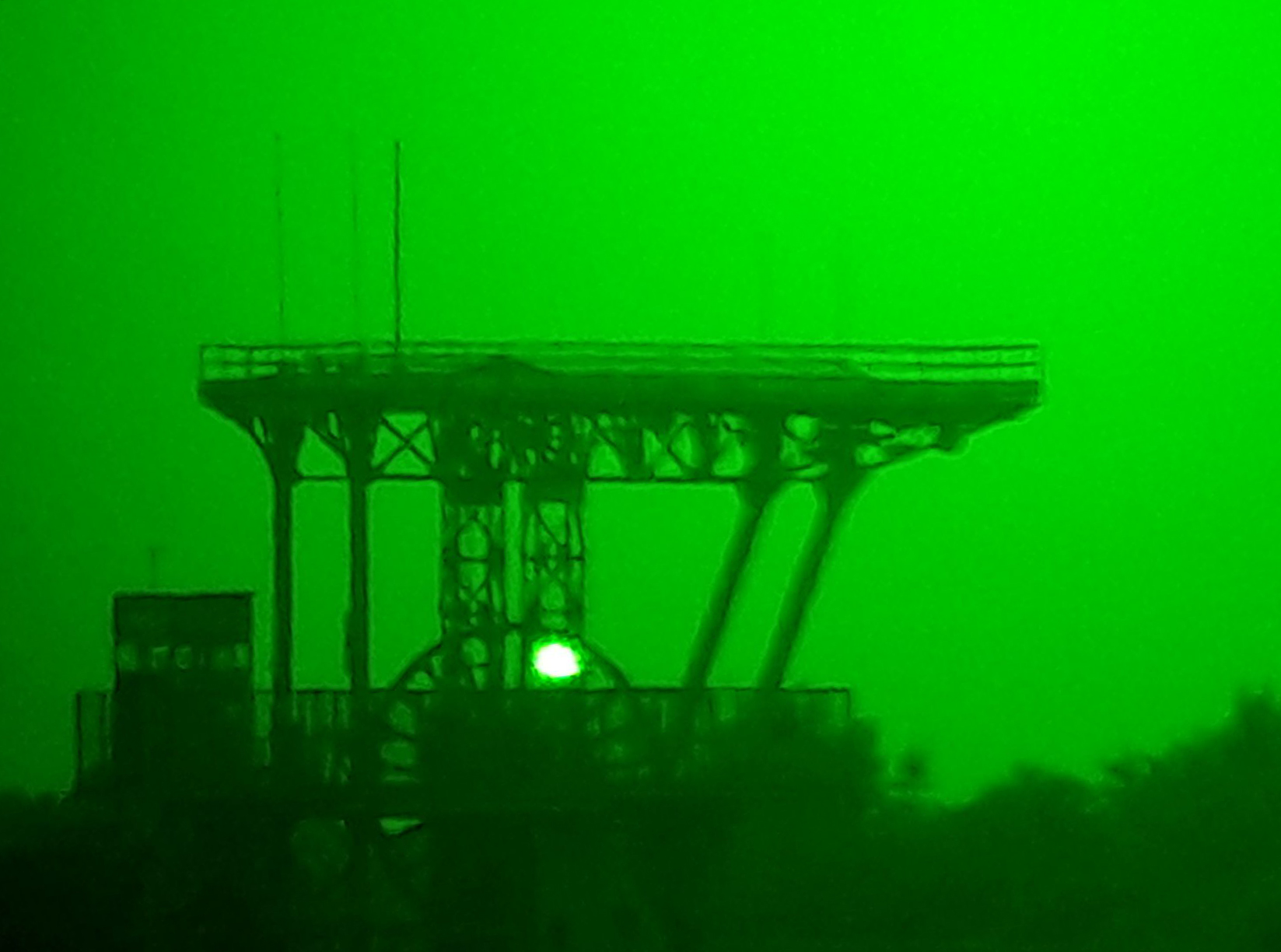} 
\caption{Left: A public exhibition for high school students on a sunny but cold day in March 2017 (the window in front of the optical device must be open to avoid image distortions and unwanted light caused by reflections). Right: the image of the spot received on the sight line with the green filter (picture taken with a smartphone).}
\label{fig4}
\end{figure}

The public was invited to measure the speed of light during an exhibition which lasted from 13$^{\textrm{th}}$ December, 2016 to 8$^{\textrm{th}}$ March, 2017 (see fig.~\ref{fig4}), and all of the 49 visits were guided by a physicist from UMONS. Among the 1~110 visitors, 462 were high school students and their teacher(s). After a slide presentation about the nature of light and the experiments of Fizeau and his successors, the principle of the measurement was described and the optical device explained. Sometimes it was necessary to realign the laser by using the micrometre screws. The presence of the sight line was very useful in this case. At the end of the explanations, some volunteers from the public had the opportunity to pilot the wheel and determine the extinction frequencies. Most of the visits took place during the day (mandatory to welcome school groups), but some were organised at night. By night, the atmosphere was very peculiar in the belfry and the ``artificial green star" was visible on the reflector to the naked eye. However, this did not make measurements easier. Moreover, it was very difficult to realign the laser in the dark. 

\begin{figure}[htb]
\includegraphics[width=9.45cm,height=5cm]{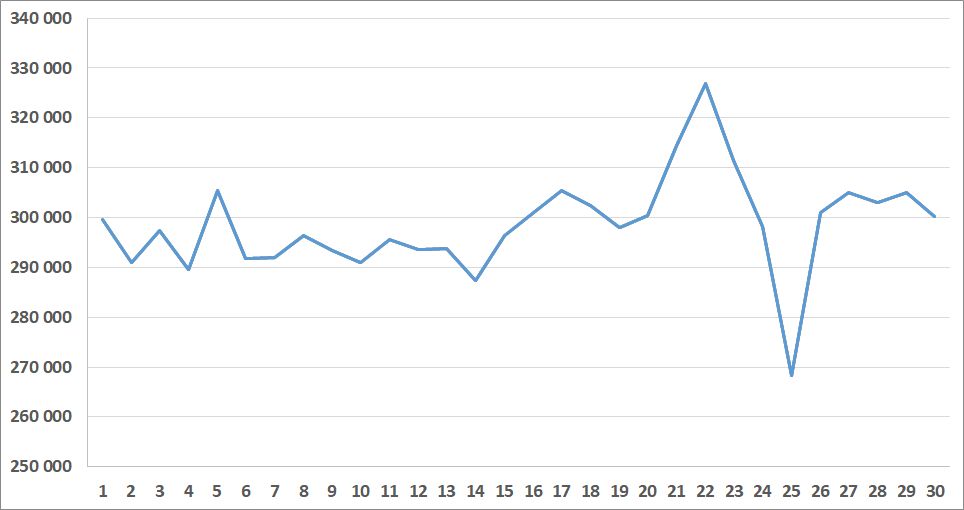} 
\caption{Values of $c_m$ in km/s found by the public, listed chronologically.}
\label{fig5}
\end{figure}

Visitors' comments were always very positive, and sometimes enthusiastic, even when it was not possible to take the measurement. Only 30 measurements were recorded with acceptable weather conditions (see fig.~\ref{fig5}). A mean value of 298~500~km/s was obtained, with a standard deviation of 9~900~km/s (3.3\%). Although accuracy was not the main goal of the experiment, the measurements were quite good, and within the range of the expected errors, except for two determinations obtained during bad weather conditions. With the two bad measurements removed, the same mean value of 298~500~km/s was obtained, with a standard deviation of 6~500~km/s (2.2\%).  

Finally, let us note an important point. It was clearly mentioned to visitors that the main purpose of the experiment was pedagogical. Since the original experiment by Fizeau, the speed of light has gained the status of a universal constant. It is now integrated into the international system of base units, where it is fixed at 299~792~458~m/s. Additionally, it was explained to the public that, from a modern point of view, what they actually measured was not the speed of light, but the distance between the chopper wheel and the reflector. 

\acknowledgments 

Many people have helped us to achieve our goal. We would like to thank Christophe Saussez and Dominique Wynsberghe from the \emph{Service Sciences et Techniques au Carr\'{e}} at UMONS for their participation in the development of the experiment, our colleagues Yves Canivez, Bjorn Maes, Michel Vou\'e and Michel Wautelet for numerous fruitful discussions about the various problems we faced, the amateur astronomers Jo\"el Bavais, Pierre Dieu, Giuseppe Monachino, Christophe Urbano and Michel Verlinden for their help in mastering imaging techniques and electronic devices, Pierre Lauginie for providing the reference of Frercks, and James Lequeux from the \emph{Observatoire de Paris} for his precious advice. Among these people, particular thanks go to Christophe Saussez for his exceptional personal investment, and Giuseppe Monachino for his idea of the reflector. We would also like to thank Manuela Valentino, the UNESCO curator for the town of Mons, and the board of the Pass for their enthusiastic collaboration. Financial support was provided by the University of Mons and the DG06 of the Public Service of Wallonia.

\end{document}